\documentclass[11pt]{article}

\usepackage[T1]{fontenc}
\usepackage{lmodern}
\usepackage[margin=1in]{geometry}
\usepackage{amsmath,amssymb,amsthm,mathtools}
\usepackage{graphicx}
\usepackage{booktabs}
\usepackage{enumitem}
\usepackage{microtype}
\usepackage[numbers,sort&compress]{natbib}
\usepackage{caption}
\usepackage{placeins}
\usepackage{xurl}
\usepackage{xcolor}
\definecolor{linkblue}{RGB}{22,61,107}
\usepackage[colorlinks=true,linkcolor=linkblue,citecolor=linkblue,urlcolor=linkblue]{hyperref}
\usepackage{cleveref}

\newtheorem{theorem}{Theorem}
\newtheorem{proposition}[theorem]{Proposition}
\newcommand{\E}{\mathbb{E}}
\newcommand{\Prb}{\mathbb{P}}
\newcommand{\stay}{\mathsf{stay}}
\newcommand{\exit}{\mathsf{exit}}

\theoremstyle{definition}
\newtheorem{definition}{Definition}

\title{Reimbursement as a Revenue Instrument in Costly Clock Auctions}
\author{
  Miriam Nelson%
  \thanks{Independent Researcher.
  Email: \texttt{miriamnelson@alumni.harvard.edu}.}
  \and
  Jack Stanick%
  \thanks{Boston College.
  Email: \texttt{stanickj@bc.edu}.}
  \and
  Edwin Lock%
  \thanks{King's College London.
  Email: \texttt{edwin.lock@kcl.ac.uk}.}
  \and
  Francisco Marmolejo-Coss\'{\i}o%
  \thanks{Boston College and IOG.
  Email: \texttt{marmolf@bc.edu}.}
}
\date{}
\hypersetup{
  pdftitle={Reimbursement as a Revenue Instrument in Costly Clock Auctions},
  pdfauthor={Miriam Nelson, Jack Stanick, Edwin Lock, Francisco Marmolejo-Cossio},
  pdfsubject={Auction design with costly continuation and reimbursement}
}

\begin{document}
\maketitle

\begin{abstract}
Bidders in an ascending auction may incur a cost each time they choose to continue, for example when bidding requires a blockchain transaction. We therefore investigate whether a seller can increase net revenue by reimbursing these costs, and how reimbursements should be distributed across auction rounds. We study auctions in which bidders must actively confirm their participation as prices rise and the seller commits to reimbursements before bidding begins. We first show how a linear program can find the cheapest payments that keep bidders’ strategies optimal when participation costs are introduced. Maximizing net revenue, however, can require changing those strategies. In a two-bidder example, we identify conditions under which reimbursing only the last scheduled decision earns the most net revenue among the schedules we analyze. This policy outperforms no reimbursement, full reimbursement, and the cheapest policy that maintains the original strategies. Some auctions end earlier than under full reimbursement, but the seller’s savings exceed the resulting loss in sale receipts. Experiments with learning bidders reproduce these revenue comparisons. In a four-bidder example, they also identify a policy with partial reimbursements in earlier rounds that earns more than reimbursing only the final decision. These results show how the seller’s preferred reimbursement timing can change across auction environments. Finally, we explain how requiring bidders to confirm continued participation limits the prices they can be charged when messages are missing, and discuss the requirements for implementing these auctions on a blockchain.

\end{abstract}

\medskip\noindent\textbf{Keywords:} blockchain auctions; mechanism design; transaction fees; bidder reimbursement; participation costs; regret minimization.

\section{Introduction}
\label{section:introduction}

% Paragraph 1. The problem and research question. Can the seller earn more by subsidizing continued bidding?
In an ascending-clock auction, the price rises from one round to the next only if at least two bidders choose to stay. When bidding costs time or money, a bidder may drop out at a lower price than she otherwise would. To keep bidders competing, the seller can offer to cover part of that cost. However, if the reimbursements exceed the resulting increase in the sale price, the seller earns less. We therefore ask:

\begin{quote}
    \emph{Can an auctioneer earn more by paying bidders to keep bidding and, if so, in which rounds should bidders be subsidized?}
\end{quote}

% Paragraph 2. Market motivation. Bidder time in web-based markets, transaction fees in blockchain markets, and the timing of subsidies.
In web-based markets, for instance, bidders in live auctions without automatic bidding can spend hours monitoring the price and responding to competing bids \citep{guan2026hassle}. In blockchain markets, a bidder may also pay for each response. For example, a clock auction run by an Ethereum smart contract can require a signed transaction at each posted price, so each confirmation costs the bidder a network fee \citep{eip1559}. Before bidding begins, the seller can use the auction's smart contract to commit to reimbursing all or part of the network fee in selected rounds. A reimbursement promised late in the auction can therefore encourage bidders to stay through earlier rounds at their own expense.

% Paragraph 3. The reimbursement rule. Who is paid, when, and how payments affect net revenue.
We study a seller who commits to a reimbursement schedule before bidding begins. Each time a bidder chooses to stay, she incurs a participation cost and earns the promised reimbursement, even if the auction ends in that round. The schedule can cover any fraction of the cost, depending on the bidder's identity, the current price, and who remained active in earlier rounds. The seller's net revenue is the sale price less the reimbursements paid to all bidders.

% Paragraph 4. The auction model. What bidders observe, when they exit, and how the object is allocated.
The seller auctions one object on a clock with a fixed sequence of increasing prices, one per round. Each bidder observes a private signal about the object's value and can learn from the prices at which others drop out. At each price before the final one, active bidders simultaneously choose whether to stay or exit. If exactly one bidder chooses to continue, she wins at the posted price. If two or more bidders continue, they advance to the next price. If everyone exits together, the seller draws a winner uniformly from those active at the start of that round and charges the posted price. If the clock reaches its final price, the same lottery applies among the bidders who reached it.

% Paragraph 5. Why the choice matters. Restoring all bidding can cost more than the extra rounds earn.
As a benchmark, suppose the seller reimburses every remaining bidder for the full cost of continuing in each round. Each bidder then faces the same payoffs as when participation is free. However, some bidders would still choose to continue if they had to pay part of the cost themselves. The seller may therefore be able to preserve a particular equilibrium of the auction with free participation while spending less than under full reimbursement. But even the cheapest rule that keeps bidders competing as long as they would in that equilibrium can cost more than the extra rounds add to the sale price.

% Paragraph 6. The two design problems. Preserve a given equilibrium at minimum cost or allow behavior to change to raise net revenue.
To study these questions, we formalize two design problems. In \emph{equilibrium preservation}, the seller starts with a particular equilibrium of the auction without participation costs. The seller seeks to minimize expected reimbursements while ensuring that the same bidding strategies and beliefs still form an equilibrium when participation is costly. Every rule that preserves this equilibrium yields the same distribution of sale prices when bidders follow it. The seller can therefore compare these rules by their reimbursement bills alone. In \emph{revenue design}, the seller chooses a reimbursement rule to maximize expected net revenue. Each rule can induce different bidding behavior, so we evaluate both the sale price and the reimbursements at the equilibrium specified for that rule.

\subsection{Our results}
\label{section:introduction-results}

% Paragraph 7. Results on preservation. The linear program and when bidders require full reimbursement.
\paragraph{Preserving an equilibrium.}
For any finite clock in our model, we construct a linear program that minimizes the expected cost of preserving a specified perfect Bayesian equilibrium of the auction without participation costs. In \cref{sec:preservation}, we prove that its feasible solutions are exactly the public reimbursement rules that preserve that equilibrium. The constraints include payments a bidder could obtain by deviating, even when the seller would never make those payments in equilibrium. For a bidder who continues, we show that her expected unreimbursed costs from that decision onward cannot exceed her gain over immediate exit when participation is free. If she is indifferent, the seller must cover her cost now and whenever she continues at a later history she can reach by following her equilibrium strategy. We apply this bound to the binary-signal example in \cref{sec:revenue}. We show that, for any finite number of bidders in this example, preserving the specified equilibrium permits savings only in the opening round. With three possible signals, we construct an example in which the least-cost payment falls when bidders' decisions reveal new information and rises again in later rounds.

% Paragraph 8. Results on revenue design. When paying only in the final decision round is optimal.
\paragraph{Choosing which rounds to reimburse.}
We solve the revenue problem for two bidders with independent binary signals drawn from the same distribution. In units where the signals are zero and one, the common value is their sum and the clock rises from zero to two in half-unit increments. We compare schedules that offer the same reimbursement to both bidders in each round, evaluated at symmetric equilibria in which high-signal bidders continue and low-signal bidders exit at the opening. In our frictionless benchmark, when both bidders have high signals, both continue to the final price. For a sufficiently small positive cost common to all bidders and rounds, we prove that the unique revenue-maximizing schedule in this comparison pays the full cost in the last decision round and nothing earlier. It earns strictly more net revenue than no reimbursement, full reimbursement, or the cheapest rule that preserves this benchmark equilibrium. When both bidders have high signals, the promised final payment keeps both bidding at their own expense in the second round. Some of these auctions then end in the third round. In our numerical example, expected sale receipts fall by 18\% relative to least-cost preservation, but the savings on reimbursements raise expected net revenue by 8\%.

% Paragraph 9. Computational results. Tests of the two-bidder prediction and a four-bidder schedule search.
\paragraph{Learning experiments.}
To test the revenue predictions, we train bidders by symmetric regret minimization and evaluate each bidder's best response exactly. Across auctions with two, three, and four bidders, 141 of the 150 strategy profiles we evaluate match the predicted probabilities of continuing to within five percentage points along the equilibrium path. With two bidders, reimbursing only the final decision round earns the most of the four benchmark rules at every cost and random seed we test. We also compare 625 schedules in a four-bidder auction with higher participation costs and larger price increments. We find that paying half the cost in the second round and a quarter in the third, with no other payments, earns about 4\% more net revenue on average than either paying only in the final decision round or offering no reimbursement.

% Paragraph 10. On-chain implementation. How fresh confirmations limit the price a bidder can be charged.
\paragraph{On-chain implementation.}
We prove that a winner's purchase price is at most the price posted one round after her last accepted confirmation. A missing or late confirmation rules a bidder out of later rounds, but she can still win at the current price if everyone exits. A bidder with no accepted confirmation can therefore win only in the opening lottery, at the opening price. We identify what an on-chain implementation requires to reproduce the auction, including timely inclusion of messages, binding choices kept private until a common release, and enforceable delivery of the object and payment of reimbursements.

\subsection{Related work}
\label{section:alternative-related-work}

% Paragraph 11. Related work on entry costs, bid preparation, and repeated opportunities to bid.
Classic models of costly entry study how the cost of submitting an offer affects who participates in an auction \citep{samuelson1985entry,levin1994equilibrium}. Gal, Landsberger, and Nemirovski show that a procurer can benefit from reimbursing part of the cost of preparing an offer \citep{gal2007participation}. In their model, reimbursement changes which suppliers submit offers. McAdams studies a second-price auction with several opportunities to submit a costly bid and shows how an observable early bid can deter later entry \citep{mcadams2015benefits}. Our seller faces a sequence of decisions about bidders who have already entered. Reimbursing bidders at one price changes the value of continuing at earlier prices.

Guan and Yuang \citep{guan2026hassle} study the recurring hassle costs of bidding in online ascending auctions. In their private-value model, these costs induce bidders to exit earlier, and their counterfactual analysis examines how higher reserve prices can mitigate the resulting revenue loss. We study the seller’s choice of reimbursement payments across rounds, accounting for how a payment changes both continuation incentives and the seller’s net revenue.

% Paragraph 12. Related work on wallet auctions and the choice of price levels.
Our two-bidder example draws on the wallet game, in which the object's common value is the sum of the bidders' private signals \citep{klemperer1998wallet,bulow2002prices}. Gon\c{c}alves and Ray study how a seller should choose discrete price levels in a Japanese--English auction for this game \citep{goncalves2024revenue}. We hold the prices fixed and choose what bidders pay to remain active at each one. 
%The rule for simultaneous exit matters for this comparison. A bidder who exits alongside her rival still has a chance to win at the current price, and that chance affects how much she stands to gain by staying.
The settlement convention also differs. In their model, when both bidders exit at a new price, the lottery winner pays the preceding price; in ours, the lottery winner pays the currently posted price. Both models charge the current price when exactly one bidder stays. The payoff from exiting alongside a rival therefore depends on the settlement convention, which affects the incentive to continue.

% Paragraph 13. Related work on learning in auctions and timely inclusion of blockchain transactions.
Our computational analysis follows work by d'Eon, Newman, and Leyton-Brown on learning bidder behavior in iterative auctions \citep{deon2024iterative}. We evaluate each learned strategy profile by computing how much a bidder could gain by deviating. On the implementation side, work on censorship in blockchain auctions explains why a bidder's ability to submit a message and have it recorded before a deadline must be part of the auction design \citep{fox2023censorship}. In our clock, missing that deadline ends the bidder's eligibility for later rounds.

\subsection{Organization}
\label{section:alternative-organization}

% Paragraph 14. Organization of the paper.
\Cref{section:model} defines the auction model. We characterize the least-cost rules that preserve a given equilibrium in \cref{sec:preservation}. In \cref{sec:revenue}, we solve the two-bidder revenue problem and derive least-cost preserving rules for examples with more bidders or richer signals. \Cref{sec:experiments} reports the learning experiments, and \cref{sec:implementation} examines on-chain implementation and bidder protection. The discussion is in \cref{sec:conclusion}. The appendices contain supplementary proofs (Appendix~\ref{app:supplementary}) and results on continuous signal cutoffs and finer price grids (Appendix~\ref{app:limits}).

\section{Model}
\label{section:model}

\subsection{The Auction}
\label{section:round-outcomes}

A seller auctions a single object to $n \geq 2$ bidders $N = \{1, \ldots, n\}$ in at most $T + 1$ rounds, numbered $0, \ldots, T$, with $T \geq 1$. The auction follows a fixed sequence of prices $0 \leq p_0 < \cdots < p_T$, with one price per round.

Write $A_t$ for the active bidder set at the start of round $t$, with $A_0 = N$. In each round $t$, the auctioneer posts price $p_t$. In each \emph{decision round} $t < T$, the active bidders simultaneously choose whether to stay or exit. A bidder who exits cannot participate in later rounds. Let $B_t \subseteq A_t$ be the set of bidders who make a stay decision (or simply a \emph{stay}). The outcome depends on how many bidders stay.
\begin{enumerate}
    \item If exactly one bidder stays, that bidder wins the object and pays $p_t$, and the auction ends.
    \item If at least two bidders stay, the auction proceeds to round $t + 1$ with active bidder set $A_{t+1} = B_t$. If this is the final round $T$, a bidder chosen uniformly at random from $A_T$ wins the object and pays $p_T$. The auction then ends.
    \item If all bidders exit, the auction ends with an \emph{all-exit lottery}.\footnote{The all-exit lottery is how a standard Japanese auction settles a tie.} A bidder chosen uniformly at random from $A_t$ wins the object and pays $p_t$.
\end{enumerate}

A \emph{public history} records the active bidder sets observed so far. At the start of decision round $t$, it is $h = (A_0, \ldots, A_t)$. We write $A(h) = A_t$ for the current active bidder set.

Let $\mathcal H$ be the finite set of decision histories. We write $h' \preceq h$, or $h \succeq h'$, when $h'$ is a prefix of $h$, including $h' = h$. Keeping the full sequence records when each bidder exited, which can affect beliefs and reimbursements even when the current active bidder set is the same.

We resolve simultaneous exit by a uniform lottery among the bidders active at the start of the round, charging the currently posted price. Thus both a sole-stayer outcome and an all-exit lottery settle at \(p_t\).

\subsection{What bidders know and pay}
\label{section:information-and-payoffs}

Bidder $i$ is risk neutral and privately observes a signal $s_i$ in a finite set $S_i$. The signal profile $s = (s_1, \ldots, s_n)$ is drawn from a joint distribution $\mu$ on $S_1 \times \cdots \times S_n$. Each $S_i$ consists of the signals bidder $i$ receives with positive probability under $\mu$. The distribution $\mu$ is common knowledge, and signals may be correlated. Bidder $i$ values the object at $v_i(s) \in [0, p_T]$, so her value may depend on other bidders' signals as well as her own. We assume that the valuation functions $v_i$ are common knowledge.

If bidder $i$ is active at a decision history $h$, her \emph{observation} is the pair $(h, s_i)$ of the public history and her private signal.

A bidder $i$ who opts to stay in a given round with history $h$ incurs a participation cost $c_i(h) \geq 0$ and earns a reimbursement $r_i(h) \geq 0$ from the seller. The participation cost could be a transaction fee for submitting the stay to a smart contract. The seller receives no revenue from participation costs.

The participation cost schedule is common knowledge. Before the auction, the seller publicly commits to a \emph{public-history policy} that sets each reimbursement using only the bidder's identity and the public history. We require the net cost $k_i(h) = c_i(h) - r_i(h)$ to be nonnegative. A bidder who exits or loses still bears her accumulated net stay costs.

Payoffs are undiscounted. A terminal outcome $z$ records the history $h$ in the last decision round $t$, the set of stayers $B_t$, the winner, and the sale price $P(z)$. Write $X_i(z) = 1$ if bidder $i$ wins and $X_i(z) = 0$ otherwise. Bidder $i$'s utility is
\begin{equation}
    u_i(s, z) = X_i(z) \bigl(v_i(s) - P(z)\bigr)
    - \sum_{\mathclap{\substack{h' \preceq h \\ i \text{ stays at } h'}}} k_i(h').
    \label{equation:bidder-utility}
\end{equation}
The seller's net revenue is
\begin{equation}
    \Pi(z) = P(z)
    - \sum_{h' \preceq h} \sum_{\mathrlap{i \text{ stays at } h'}} r_i(h').
    \label{equation:seller-revenue}
\end{equation}

A higher sale price benefits the seller only to the extent that the increase exceeds any additional reimbursements needed to induce that price.

\subsection{Equilibrium and the seller's choice}
\label{section:equilibrium-and-design}

A \emph{behavioral strategy} $\sigma_i$ specifies the probability $\sigma_i(h, s_i)$ that bidder $i$ chooses to stay given her observation $(h, s_i)$. Her \emph{beliefs} $\beta_i$ assign to each possible observation a probability distribution $\beta_i(\cdot \mid h, s_i)$ over the other bidders' signal profiles that have positive probability under $\mu$ given $s_i$. An \emph{assessment} $(\sigma, \beta)$ specifies a strategy $\sigma_i$ and beliefs $\beta_i$ for each bidder $i$.

The prior $\mu$, the bidders' strategies $\sigma$, and the allocation lotteries determine a joint distribution over signal profiles $s$ and terminal outcomes $z$. Ex ante expected utility and expected net revenue are the averages of the payoffs in \eqref{equation:bidder-utility} and \eqref{equation:seller-revenue} under this distribution.

After an observation $(h, s_i)$, bidder $i$ evaluates her expected utility by averaging $u_i(s, z)$ over the other bidders' signals according to $\beta_i(\cdot \mid h, s_i)$, the actions from the current round onward under the strategies being evaluated, and the allocation lotteries.

A \emph{perfect Bayesian equilibrium} (PBE), in the weak sense used here, is an assessment $(\sigma, \beta)$ that satisfies two conditions.
\begin{enumerate}[label=(\roman*)]
    \item After any possible observation $(h, s_i)$, bidder $i$ cannot improve her expected utility by changing how she plays from that point onward, given her current beliefs and the other bidders' strategies.
    \item For each observation that occurs with positive probability under the prior $\mu$ and the strategies $\sigma$, the bidder's beliefs follow by conditioning on that observation using Bayes' rule.
\end{enumerate}

Let $c$ and $r$ denote the schedules of participation costs and reimbursements, and write $\Gamma(c, r)$ for the resulting game. The \emph{frictionless game} $\Gamma^0 = \Gamma(0, 0)$ is the same auction with no participation costs and no reimbursements. The winner still pays the sale price.

\begin{definition}[Equilibrium preservation]
\label{definition:equilibrium-preservation}
Fix participation costs $c$ and a target PBE $(\sigma^0, \beta^0)$ of $\Gamma^0$. A reimbursement rule $r$ \emph{preserves} the target if the same assessment is a PBE of $\Gamma(c, r)$.
\end{definition}

A reimbursement promised at an \emph{off-path history}, one the target never reaches, costs the seller nothing under the target. It can still affect a bidder's incentives if a deviation reaches that history.

A rule that preserves the target can also support other equilibria, so we evaluate each rule at a specified equilibrium. At the target assessment, all preserving rules yield the same expected sale receipts. Among those rules, the seller maximizes expected net revenue by minimizing expected reimbursements. In the revenue design problem, a reimbursement rule can change how bidders play.

Under an \emph{anonymous round-only} reimbursement rule, every bidder who stays in round $t$ receives $\rho_t$. Since $p_t$ uniquely identifies the round, the rule can also be specified by price. The preservation program allows the full class of public-history policies. The two-bidder revenue design problem is restricted to anonymous round-only rules.

\section{Preservation at minimum cost}
\label{sec:preservation}

\subsection{The preservation program}

Full reimbursement, $r_i(h)=c_i(h)$, makes every bidder's payoff identical to its frictionless payoff on every terminal path. It therefore preserves every frictionless assessment. The seller may be able to spend less if the target gives bidders enough incentive to bear some of their own stay costs.

The key simplification is to evaluate a fixed continuation plan while varying payments. Once a plan and the rivals' strategies are fixed, the probabilities of reaching later histories are fixed as well. Reimbursement then affects only the net cost paid on each path. Incentive constraints will determine which payment rules actually make the target optimal.

Given bidder $i$'s observation $\iota = (h, s_i)$, let $D_i(\iota)$ be her finite set of pure continuation plans. Each plan specifies an action now and an action for each possible later observation. Fix rivals' target strategies $\sigma_{-i}^0$ and the current target belief $\beta_i^0(\cdot\mid\iota)$. For a continuation $\tau_i$, write $U_i^{c,r}(\tau_i\mid\iota)$ for expected utility from $h$ onward, excluding costs and transfers already incurred, and $U_i^0(\tau_i\mid\iota)$ for the corresponding frictionless payoff.

Define $a_i^{\tau_i,\iota}(h')$ as the conditional probability that $h'\succeq h$ is reached and $i$ stays there under this continuation and the fixed rivals. Each such probability weights the net cost paid at that history. Thus
\begin{equation}
U_i^{c,r}(\tau_i\mid\iota)=U_i^0(\tau_i\mid\iota)
-\sum_{h'\succeq h}a_i^{\tau_i,\iota}(h')\bigl(c_i(h')-r_i(h')\bigr).
\label{eq:affine}
\end{equation}
Histories at which $i$ is inactive are omitted from the sum. This identity separates a fixed frictionless payoff from a weighted sum of net costs. In particular, the payoff is affine in the publicly promised reimbursements.

The seller needs a different probability weight for expenditure. Let $\lambda_i^0(h)$ be the unconditional probability under $\mu$ and $\sigma^0$ that $h$ is reached and $i$ stays. Whereas $a_i^{\tau_i,\iota}(h')$ follows one bidder's continuation given her observation, $\lambda_i^0(h)$ measures payment frequency under the target from the start of the auction. An off-path history can have zero weight in the objective and positive weight in a deviation comparison.

The seller minimizes target expenditure subject to making the target weakly better than every continuation deviation:
\begin{align}
\min_r\quad &\sum_{h\in\mathcal H}\sum_{i\in A(h)}\lambda_i^0(h)r_i(h) \label{lp:objective}\\
\text{s.t.}\quad &U_i^{c,r}(\sigma_i^0\mid\iota)\ge U_i^{c,r}(\tau_i\mid\iota)
&&\forall i,\iota,\tau_i\in D_i(\iota), \label{lp:ic}\\
&0\le r_i(h)\le c_i(h)&&\forall h,\ i\in A(h). \label{lp:box}
\end{align}
The first line is expected reimbursement expenditure. The second imposes sequential incentives, and the third enforces the no-overpayment restriction.

\begin{theorem}[Finite preservation program]
\label{thm:lp}
For any finite costly clock and specified frictionless target $(\sigma^0,\beta^0)$, a public rule preserves the target if and only if it is feasible for \eqref{lp:objective}--\eqref{lp:box}. This is a feasible LP and attains its minimum.
\end{theorem}

\begin{proof}
Fix a bidder's observation and a continuation plan. Couple the costly and frictionless games using the same rivals' actions, signals, and allocation lotteries. Allocations and prices coincide because the plan is fixed; only the net costs of stays differ. Taking expectations gives \eqref{eq:affine}. The identity also applies to the possibly mixed target by averaging over its action randomizations.

A behavioral continuation in this finite perfect-recall game is a lottery over pure continuation plans with the same distribution of play. Hence the target is sequentially rational if and only if, for every bidder $i$ and every possible observation $\iota$, her target continuation weakly beats every plan in $D_i(\iota)$. Those are exactly the inequalities in \eqref{lp:ic}. Retaining the target strategies retains its on-path signal distributions, so $\beta^0$ remains Bayes consistent wherever required; its specified off-path beliefs are unchanged. Conversely, preserving the target requires every displayed incentive inequality.

The objective is linear, and \eqref{eq:affine} makes every incentive constraint affine in $r$. Full reimbursement is feasible because it restores each frictionless payoff. Finally, the feasible set is a closed subset of the compact reimbursement box in \eqref{lp:box}, so the objective attains its minimum.
\end{proof}

The program chooses off-path promises together with payments the seller expects to make. It uses complete continuation plans because a deviation can change both future participation and the public information a bidder subsequently observes. Under an additional consistency condition on beliefs, it suffices to check each current action followed by a return to the target.\footnote{Beliefs are \emph{deviation-consistent} if, whenever a bidder's later observation has positive probability under one of her unilateral continuation deviations, her assigned belief given that observation agrees with updating her current belief through the deviation and her rivals' target play. Appendix~\ref{app:preservation} proves the resulting local deviation reduction. The all-plan LP does not require this condition.} This can shorten the constraint list. The result is nevertheless a formulation for a finite game; an exponentially large game tree can still produce an exponentially large program.

\subsection{Incentive margins as cost budgets}

The program gives a complete test for preservation. A particularly useful necessary constraint comes from comparing a target stay with immediate exit.

Write $\stay\to\sigma_i^0$ for staying now and then following the target at later decisions. At a target stay, define the frictionless continuation margin
\begin{equation}
W_i^0(h,s_i):=U_i^0(\stay\to\sigma_i^0\mid h,s_i)-U_i^0(\exit\mid h,s_i).
\label{eq:margin}
\end{equation}
This is the gain from the stay-first continuation before participation costs are introduced. The exit payoff includes the current all-exit lottery and need not be zero. Thus a margin compares two full continuation payoffs, rather than identifying the stay payoff with the margin.

\begin{proposition}[Zero margins force exact reimbursement]
\label{prop:zero}
Suppose that under the target, bidder $i$ stays with positive probability given $(h, s_i)$. Under every preserving rule, the expected net cost incurred by staying now and returning to the target is at most $W_i^0(h,s_i)$. If this margin is zero, then $r_i(h)=c_i(h)$, and every later stay reached with positive probability on that continuation is also reimbursed exactly.
\end{proposition}

\begin{proof}
Write $\iota = (h, s_i)$ for bidder $i$'s observation and let $\theta > 0$ be her target stay probability given $\iota$. The target payoff is a mixture of the stay-first continuation and immediate exit. If the former were worse than exit, putting positive weight on it would contradict optimality of the target. Hence the stay-first continuation is itself weakly better than exit under every preserving rule.

Exit produces no current or future own stay costs. Applying \eqref{eq:affine}, the expected net cost of the stay-first continuation must therefore satisfy
\[
0\le K_i(h,s_i):=\sum_{h'\succeq h}a_i^{\stay\to\sigma_i^0,\iota}(h')k_i(h')\le W_i^0(h,s_i).
\]
Every coefficient and every $k_i(h')$ is nonnegative, and the coefficient on the current stay is one. If the margin is zero, each positive-probability term must vanish. In particular, the current net cost is zero, and so are all later net costs reached with positive probability on this continuation.
\end{proof}

The restriction rules out offsetting one unreimbursed stay with a subsidy larger than another stay's cost: overpayment is unavailable. A bidder with no frictionless advantage from continuing cannot absorb any positive expected net cost along that continuation. Because reimbursement is public, a payment forced by the weakest type at a history is also paid to stronger types who would stay for less. We call this \emph{inframarginal leakage}.

A positive margin instead supplies a budget for the expected sum of net continuation costs. Suppose the margin is $0.10$. A current net cost of $0.04$ and a later net cost of $0.08$ incurred with probability one half consume $0.08$ of that budget. If both costs are incurred for certain, they consume $0.12$ and violate the constraint. The same margin cannot be spent independently at every future stay. Passing this one budget test need not satisfy all the LP's other incentive comparisons.

The budget interpretation can also give sufficient bounds when margins are uniformly positive. For example, a uniform margin $w>0$ and at most $D$ remaining own stays allow net cost at most $w/D$ per stay for a pure target with deviation-consistent beliefs, provided these conditions also hold for observations that never occur under the target. Appendix~\ref{app:preservation} supplies the backward argument. A continuous cutoff with staying types arbitrarily close to indifference need not have such uniform slack.

The cost of preserving an equilibrium depends on when bidders have sufficient incentives to continue without reimbursement. Consequently, even when participation costs are constant across rounds, the cheapest preserving policy need not reimburse every round at the same rate. The examples in Section~\ref{sec:extensions} illustrate how information revealed during the auction can create new opportunities to reduce payments.

More fundamentally, preserving the original equilibrium need not be the seller's most profitable choice. A reimbursement policy can change when bidders exit, affecting both the final sale price and the payments made along the way. We turn to this revenue-design problem next.

\section{Reimbursement as revenue design}
\label{sec:revenue}

We first study a small common-value environment in which both parts of this calculation can be carried out exactly. Three benchmarks show what is bought by no reimbursement, by full reimbursement, and by the cheapest preserving rule. We then optimize payment timing over all anonymous price-dependent rules on a specified equilibrium branch.

\subsection{Binary wallet environment}

There are two bidders with independent signals $s_i\in\{0,a\}$, where $a>0$ and $\Prb(s_i=a)=q\in(0,1)$. Signal zero is \emph{low} and signal $a$ is \emph{high}. Both bidders value the object at the common wallet value $V=s_1+s_2$. Thus the possible values are $0$, $a$, and $2a$, according to whether zero, one, or both signals are high. Decision prices are $0,a/2,a,3a/2$, followed by terminal cap $2a$. Each stay costs $c>0$, and $\kappa:=c/a$ is this cost measured relative to the signal size.

The opening at price zero serves an informational role. In an \emph{opening-sorting} assessment, low types exit and high types stay. If just one bidder stays, that bidder wins immediately at zero. If both stay, each can infer that its rival is high. Their common value is then known to be $2a$, and the remaining three decision rounds form a complete-information continuation. We call this continuation the \emph{war}: each bidder would like the rival to exit first, but staying may itself be costly.

\paragraph{The frictionless target.}
When there are no stay costs, consider the assessment in which both revealed high types stay through the three war prices and reach the cap. They each obtain zero surplus at $2a$. Exiting earlier also gives zero, because the rival stays for certain. Thus every target stay in the war has zero frictionless stay-minus-exit margin.

At the opening, a high type has positive margin. Against a low rival, staying gives it the entire surplus $a$ at price zero, whereas exiting produces the all-exit lottery and an expected surplus $a/2$. Against a high rival, staying leads to the zero-surplus war and exiting gives zero because the rival stays. Averaging these two cases gives the opening margin
\begin{equation}
b:=\frac{(1-q)a}{2}>0.
\label{eq:b}
\end{equation}
This is the only positive margin along the target's high-type path.

To complete the equilibrium, we must also specify what happens if a low type unexpectedly stays at the opening. After two opening stays, each bidder believes her rival has signal $a$, even if her own signal is low. Prescribe a sequential best response for a low type who stays. A low type that enters against a high rival has true value $a$, while the rival continues to act as if the value were $2a$. The low type can exit before paying an excessive price but cannot obtain positive surplus by waiting for that rival to leave. Against a low rival it wins at price zero with value zero. Consequently, exiting is optimal for low types at the opening. These beliefs are Bayesian wherever the target reaches, so the specified assessment is a frictionless PBE.

The asymmetry between the opening and the war is central to preservation. A high type can bear some opening cost out of $b$. Its later stays have no such slack.

\subsection{Three benchmarks}
\label{sec:binary-benchmarks}

\paragraph{No reimbursement (None).}
First leave every stay unpaid. Even though the frictionless all-stay war cannot survive a positive cost, there is an opening-sorting PBE when $0<\kappa<1/4$. The war now involves mixing at some prices.

To calculate that continuation, work backward from the cap. At a current war price $p$, let $W'$ be a bidder's continuation payoff if both bidders stay and advance, excluding the current stay cost. Let $\sigma$ be the rival's probability of staying now. Conditional on the two high signals already being revealed, exit and stay give
\begin{align*}
E(\sigma)&=(1-\sigma)(2a-p)/2,\\
S(\sigma)&=-c+(1-\sigma)(2a-p)+\sigma W'.
\end{align*}
The first expression includes the all-exit lottery: an exiter receives half the available surplus only when its rival also exits. A stayer wins the entire surplus when its rival exits, and otherwise receives the next-round payoff. It pays $c$ on both branches, including an immediate win.

At $3a/2$, the next payoff is zero. Indifference between the two actions gives rival stay probability $1-4\kappa$ and a bidder payoff of $c$. At price $a$, that next-round payoff exactly covers the current cost, so both bidders stay and their payoff at $a$ is zero. At price $a/2$, the same indifference calculation then gives stay probability $1-4\kappa/3$ and payoff $c$. Thus, writing $\sigma_t$ for each active high bidder's stay probability at war price $ta/2$, $t=1,2,3$, we obtain
\begin{equation}
(\sigma_1,\sigma_2,\sigma_3)=\left(1-\frac{4\kappa}{3},1,1-4\kappa\right).
\label{eq:none-strategy}
\end{equation}
These $\sigma_t$ are action probabilities after the high signals have been revealed. They are different from $q$, the prior probability of receiving a high signal. At each war price, stay minus exit is strictly decreasing in the rival's stay probability. The displayed continuation is therefore the unique symmetric solution; at the middle price, staying is strictly preferable whenever the rival stays with probability below one.

\paragraph{Checking the opening.}
Solving the revealed war is not enough to establish an equilibrium: the opening actions must lead types into that war voluntarily. A high type's opening gain is $(1-q)(a/2-c)>0$. Against a high rival, its opening cost is exactly offset by the war entry payoff $c$; against a low rival, staying gives $a-c$ rather than the exit-lottery payoff $a/2$.

A low type that deviates at the opening pays $c$. Against a high rival, its best payoff at $a/2$ comes from exiting and equals $(1-\sigma_1)a/4=c/3$. Staying gives at most $-c+(1-\sigma_1)a/2=-c/3$, because all later prices are at least its true value $a$ and no later continuation can give it positive surplus. Against a low rival, the opening deviation wins a zero-value object at zero and still pays $c$. Its total deviation payoff is therefore $-c+qc/3<0$. The same beliefs used above, together with sequential low-type best responses, complete the sorting PBE.

\paragraph{Revenue under None.}
Only the two-high event, which has probability $q^2$, produces a positive sale price. Conditional on this event, the sale occurs at $a/2$ unless both bidders stay there. If both stay, they certainly pass price $a$, then sell at $3a/2$ unless both survive to the cap. Writing $\Pi^{\mathrm{none}}$ for ex ante seller net revenue gives
\begin{equation}
\begin{aligned}
\Pi^{\mathrm{none}}&=aq^2R_{\mathrm{none}}(\kappa),\\
R_{\mathrm{none}}(\kappa)&=\frac{1-\sigma_1^2}{2}
+\sigma_1^2\left(\frac{3}{2}+\frac{\sigma_3^2}{2}\right).
\end{aligned}
\label{eq:none-revenue}
\end{equation}
The function $R_{\mathrm{none}}$ records the conditional stopping-price calculation in units of $a$. There are no transfers to subtract under this rule.

\paragraph{Least-cost preservation and full reimbursement.}
Now return to the frictionless target. Proposition~\ref{prop:zero} forces payment $c$ on every war stay: those stays have zero margin. After removing all later costs, the high type's opening constraint is simply $b-(c-\rho_0)\ge0$. The unique least-cost payments on target-path stays are therefore
\begin{equation}
\rho^{\mathrm{pres}}=\bigl(\max\{0,c-b\},c,c,c\bigr).
\label{eq:pres-policy}
\end{equation}
These amounts are sufficient as well as necessary. Fully reimburse every post-opening stay, including off-path stays, so that continuation incentives remain frictionless. The opening cost left to a high type is at most its margin $b$, while a nonnegative opening net cost cannot make a low type's deviation more attractive. Each required payment has positive target reach-and-stay probability, proving uniqueness of the minimum amounts on target-path stays. Off-path payments need not be unique.

We call this rule Preserve and call reimbursement of every stay Full. They evaluate the same frictionless assessment, so their gross receipts coincide. The two-high event produces six post-opening stays, three by each bidder; the expected number of opening stays is $2q$. Subtracting the corresponding transfers from gross receipts $2aq^2$ gives
\begin{align}
\Pi^{\mathrm{pres}}&=2aq^2-6cq^2-2q\max\{0,c-b\}, \label{eq:pres-rev}\\
\Pi^{\mathrm{full}}&=2aq^2-6cq^2-2qc. \label{eq:full-rev}
\end{align}
Thus Preserve saves $2q\min\{c,b\}$ relative to Full, entirely at the opening. It pays for all the zero-margin war stays.

\paragraph{Comparing preservation with no reimbursement.}
The comparison with None includes a behavioral change as well as a transfer change. Preserve earns more if
\[
0<\kappa<\min\{\bar\kappa,(1-q)/2\},
\]
where $\bar\kappa\approx0.108629$ is the unique root in $(0,1/4)$ of $64\kappa^3-128\kappa^2+96\kappa-9=0$. To verify this claim, note that $\kappa<(1-q)/2$ makes the opening payment in \eqref{eq:pres-policy} zero. Subtracting \eqref{eq:none-revenue} from \eqref{eq:pres-rev} then gives
\[
\Pi^{\mathrm{pres}}-\Pi^{\mathrm{none}}
=-\frac{2aq^2\kappa}{9}h(\kappa),\quad
h(x)=64x^3-128x^2+96x-9.
\]
The derivative $h'(x)=192x^2-256x+96$ is positive everywhere, since its discriminant is negative and its leading coefficient is positive. Since $h(0)=-9$ and $h(1/4)=8$, the stated root and comparison follow.

The benchmarks separate three effects. None spends nothing and changes war behavior. Full restores the frictionless play but pays at the opening even though a high type already has margin $b$. Preserve removes that opening payment where the margin permits, while buying every frictionless war stay. The next question is whether the seller should buy all of that war participation.

\subsection{Optimal timing on the sorting branch}

An anonymous price-dependent policy is $\rho=(\rho_0,\rho_1,\rho_2,\rho_3)\in[0,c]^4$. We evaluate each policy on its \emph{symmetric sorting branch}: low types exit and high types stay at the opening, and the revealed war uses the symmetric continuation obtained by backward induction. A policy enters the comparison only if those opening actions are optimal.

This selection fixes the meaning of revenue under a policy. Sorting describes behavior, while the signals themselves remain private. Two opening stays reveal high signals because low types exit in this assessment. We first solve the continuation under that inference and then check that neither type wants to change its opening action. The comparison therefore includes policies with different war behavior, but requires that each assessed war be supported by the same type of informative opening.

\begin{theorem}[Optimal back-loading]
\label{thm:backload}
Suppose $0<\kappa<\min\{1/4,(1-q)/2\}$. Every war-payment vector $(\rho_1,\rho_2,\rho_3)\in[0,c]^3$ paired with $\rho_0=0$ admits a symmetric sorting PBE. Among anonymous price-dependent policies admitting and evaluated on this branch, the unique revenue-maximizing policy is
\begin{equation}
\widehat\rho=(0,0,0,c).
\label{eq:backload-policy}
\end{equation}
Its selected war stay probabilities are $(1,1-2\kappa,1)$ and its net revenue is
\begin{equation}
\widehat\Pi=aq^2\bigl[1+(1-2\kappa)^3\bigr].
\label{eq:backload-rev}
\end{equation}
It strictly exceeds all three benchmarks:
\begin{align}
\widehat\Pi-\Pi^{\mathrm{pres}}&=4aq^2\kappa^2(3-2\kappa)>0,\label{eq:gap-pres}\\
\widehat\Pi-\Pi^{\mathrm{none}}&=\frac{2aq^2\kappa}{9}(9-42\kappa+92\kappa^2-64\kappa^3)>0,\label{eq:gap-none}\\
\widehat\Pi-\Pi^{\mathrm{full}}&=4aq^2\kappa^2(3-2\kappa)+2qc>0.
\label{eq:gap-full}
\end{align}
\end{theorem}

The cost restrictions are $c<a/4$ and $c<b$. They ensure that the war can be analyzed in a low-cost region and that a high type has enough opening slack to enter without reimbursement. The theorem makes two claims: every war-payment vector can be supported with an unpaid sorting opening, and, among the resulting assessed revenues, a unique vector is best.

\paragraph{How the proposed rule works.}
The effect of the final payment is easiest to understand backward, starting at the last decision:
\begin{enumerate}[leftmargin=*,itemsep=3pt,topsep=3pt]
    \item At $3a/2$, reimbursement removes the last stay cost. Both high bidders stay and reach the cap $2a$, where each earns zero surplus.
    \item At $a$, the current stay is unpaid. Continuing against a rival who stays leads to the zero-payoff final round. Each bidder mixes with probability $1-2\kappa$; indifference gives it an expected payoff of $c$ at this middle price.
    \item At $a/2$, a stay costs $c$ but leads to the middle-round payoff $c$ if both remain. The continuation covers that cost, and both bidders stay.
\end{enumerate}
Some wars consequently end at $a$ instead of $2a$. The seller gives up gross sale receipts but saves more in reimbursements. The final payment also supports an earlier unpaid stay by raising the value of that earlier continuation. A payment can therefore matter at prices before the one at which it is made.

This explanation establishes what the proposed rule does. It does not yet compare the rule with every other payment vector. Section~\ref{sec:binary-proof} supplies that optimization argument after a numerical illustration.

\paragraph{A numerical illustration.}
Set $a=1$, $q=0.3$, and $c=0.1$. The opening margin is $b=0.35$, so the high type can bear the entire opening cost. Table~\ref{tab:worked-example} evaluates the four policies from their formulas; these are analytical calculations, separate from the learning experiments.

\begin{table}[htbp]
\centering
\caption{Binary example at $a=1$, $q=0.3$, $c=0.1$. Payments are expressed as fractions of $c$. War stays are conditional on two high opening signals; net revenues are rounded to four decimals.}
\label{tab:worked-example}
\begin{tabular}{@{}lccc@{}}
\toprule
Policy & $\rho/c$ & War stays & Net revenue\\
\midrule
None & $(0,0,0,0)$ & $(13/15,1,3/5)$ & $0.1248$\\
Preserve & $(0,1,1,1)$ & $(1,1,1)$ & $0.1260$\\
Full & $(1,1,1,1)$ & $(1,1,1)$ & $0.0660$\\
Back-loaded & $(0,0,0,1)$ & $(1,4/5,1)$ & $0.1361$\\
\bottomrule
\end{tabular}
\end{table}

Preservation earns gross revenue $0.18$ and pays $0.054$, leaving $0.126$. Back-loading earns less gross revenue, $0.1476$, but pays only $0.01152$, leaving $0.13608$. Its transfer saving therefore exceeds its loss in sale receipts. Full reimbursement induces the same play as preservation but additionally pays $2qc=0.06$ at the opening, explaining its much lower net revenue.

The policy also purchases fewer expected stays than preservation and changes the distribution of stopping prices. Here \emph{timing} means which actions receive reimbursement; transfers paid for fixed behavior are not discounted. The optimum remains conditional on the comparison class: Theorem~\ref{thm:backload} does not eliminate asymmetric continuations or pooling opening equilibria, and does not optimize over arbitrary history-dependent policies.

\subsection{Proof of optimal back-loading}
\label{sec:binary-proof}

The proof follows the order in which an earlier decision depends on later play. First, we solve the symmetric war for any payment vector. Second, we optimize the last two payments while holding fixed the payoff they deliver to a bidder at the middle price. Third, we choose the first war payment and the desired middle-price payoff. Finally, we check that the opening really sorts types and choose its payment. This separates the optimization within the revealed war from the incentives that make the war's revelation valid.

\begin{proof}[Proof of Theorem~\ref{thm:backload}]
Put $C=\kappa$ and $x_t=(c-\rho_t)/a\in[0,C]$ for $t=0,1,2,3$. An $x_t$ is a cost left with the bidder: $x_t=C$ means no reimbursement, and $x_t=0$ means full reimbursement. Let $W_t$ and $R_t$ be bidder continuation utility and seller net revenue from war round $t$ onward, divided by $a$ and conditional on two high bidders remaining. Thus $W_4=0$ and $R_4=2$ at the terminal cap.

Throughout this proof, $C,x_t,W_t,R_t$ are normalized by $a$. Named prices such as $a/2$, original payments $\rho_t$, and ex ante revenue $\Pi$ retain monetary units. Also, $W_t$ is a continuation payoff level under the candidate rule; it differs from $W_i^0(h,s_i)$ in Section~\ref{sec:preservation}, which is a frictionless stay-minus-exit margin.

\paragraph{Step 1: solve the symmetric continuation.}
At war round $t$, put $\zeta_t=(2-t/2)/2$, half the remaining gross surplus in normalized units. If the bidder exits, it receives $\zeta_t$ only when its rival also exits. If it stays, it wins the entire remaining surplus when the rival exits and otherwise obtains $W_{t+1}$. Subtracting the current net cost gives the stay-minus-exit difference against rival stay probability $\sigma$:
\[
D_t(\sigma)=(1-\sigma)\zeta_t+\sigma W_{t+1}-x_t.
\]
The first term is the gain from replacing a half-share lottery with a certain win; the second is the value of reaching the next price; the last is the cost paid for staying now.

Inductively, $0\le W_{t+1}\le C<\zeta_t$, so $D_t$ is strictly decreasing and $D_t(0)>0$. There are therefore two cases. If the next-round payoff covers the current cost, staying is optimal even against a rival who stays for certain. Otherwise the symmetric continuation requires a mixture that makes both actions equally attractive. The unique symmetric solution is
\begin{equation}
(\sigma_t,W_t)=
\begin{cases}
(1,W_{t+1}-x_t),&x_t\le W_{t+1},\\[2pt]
\left(\dfrac{\zeta_t-x_t}{\zeta_t-W_{t+1}},
(1-\sigma_t)\zeta_t\right),&x_t>W_{t+1}.
\end{cases}
\label{eq:war-recursion}
\end{equation}
The equality case has $\sigma_t=1$. In the mixed case, the bidder's payoff equals its exit payoff, and indifference also gives $W_t=x_t-\sigma_tW_{t+1}\le C$. In the certain-stay case, $0\le W_t=W_{t+1}-x_t\le C$. This completes the induction, starting from $W_4=0$.

Seller revenue follows the same stopping events. The sale occurs now unless both bidders stay, and otherwise the seller carries forward $R_{t+1}$. A current stayer receives reimbursement even when it wins immediately. Consequently,
\begin{equation}
R_t=(1-\sigma_t^2)t/2-2\sigma_t(C-x_t)+\sigma_t^2R_{t+1}.
\label{eq:seller-recursion}
\end{equation}
There are $2\sigma_t$ expected current stayers, including immediate winners. Current transfers therefore use weight $2\sigma_t$, whereas future revenue uses $\sigma_t^2$, the probability that both bidders stay. Keeping these weights separate is essential: paying for a stay and reaching the next round are different events.

For the last war price, the formulas reduce to
\begin{align*}
W_3&=x_3,\qquad \sigma_3=1-4x_3,\\
R_3&=2-2C+(8C-2)x_3.
\end{align*}
These formulas include $x_3=0$, where both bidders stay and their continuation payoff is zero.

\paragraph{Step 2: optimize the last two payments for a fixed middle-price payoff.}
The first war round needs only two pieces of information about later play: the bidder payoff $W_2$ that supports staying, and the seller revenue $R_2$ produced by that play. Write $w=W_2\in[0,C]$. We now hold $w$ fixed and find the greatest possible $R_2$. Denote this revenue frontier by $F(w)$.

We must consider both branches of \eqref{eq:war-recursion}. On the branch $x_2\le x_3$, both bidders stay at round 2, so $w=x_3-x_2$ and
\[
R_2=2-4C+8Cx_3-2w
\le F_A(w):=2-4C+8C^2-2w.
\]
For a fixed promised payoff $w$, revenue increases with $x_3$. The upper bound is attained uniquely at $x_3=C$ and $x_2=C-w$. This describes the best rule among those that keep both bidders for certain at the middle price.

On the mixed branch $x_2>x_3$, the same recursion gives
\begin{align*}
w&=\frac{x_2-x_3}{1-2x_3},\qquad \sigma_2=1-2w,\\
x_2&=x_3+w(1-2x_3).
\end{align*}
Thus a fixed $w$ also fixes the middle-price stay probability. The remaining design freedom is how the net costs at the middle and final prices deliver that payoff. For fixed $w>0$, feasibility gives $0\le x_3\le(C-w)/(1-2w)$. Substituting in \eqref{eq:seller-recursion} yields
\[
\left.\frac{\partial R_2}{\partial x_3}\right|_w
=8C(1-2w)^2>0.
\]
Revenue is therefore maximized by raising $x_3$ as far as possible while maintaining $w$. The mixed-branch optimum has $x_2=C$ and $x_3=(C-w)/(1-2w)$, with revenue
\begin{equation}
F(w)=F_A(w)+4Cw(1-4C+2w).
\label{eq:frontier}
\end{equation}
For $w>0$, the added term is strictly positive because $C<1/4$. Allowing a middle-price mixture consequently improves on the best certain-stay rule delivering the same bidder payoff. At $w=0$, the optimum is uniquely attained at $x_2=x_3=C$, and \eqref{eq:frontier} extends to this endpoint. We have therefore optimized over both branches, with a unique best pair of later net costs for each $w$.

\paragraph{Step 3: optimize the first war payment and the payoff to deliver.}
We next choose $x_1$ and $w$. The first war round's stay probability is strictly positive, so increasing $R_2$ at fixed $w$ increases $R_1$. We can therefore restrict attention to the frontier $F(w)$ without losing any candidate optimum.

If $x_1\le w$, both bidders stay for certain. Revenue is $F(w)-2(C-x_1)$, uniquely maximized at $x_1=w$. Lowering the current net cost below $w$ would spend more on reimbursement without changing current survival. Define the best revenue on this branch by
\[
G(w):=F(w)-2(C-w),\qquad
G'(w)=4C(1-4C+4w)>0.
\]
Within the certain-stay branch, the seller therefore prefers the largest feasible payoff $w=C$.

It remains to rule out a different possibility: charging $x_1>w$ induces mixing at the first war price and might save enough reimbursement to improve revenue. For this case, write $\alpha=(3/4-x_1)/(3/4-w)$ for the resulting stay probability. The largest revenue on the mixed branch is
\[
H(w,x_1)=\frac{1-\alpha^2}{2}+\alpha^2F(w)-2\alpha(C-x_1).
\]
We compare this expression with $G(w)$ by varying $x_1$ while keeping $w$ fixed. For the boundary derivative, define the polynomial
\begin{align*}
N(C,w)={}&-64C^2w+32C^2+32Cw^2\\
&+16Cw-20C+3.
\end{align*}
Direct differentiation gives
\begin{align*}
H_{x_1x_1}(w,x_1)
&=\frac{128C(2w-1)(1+w-2C)}{(4w-3)^2}<0,\\
H_{x_1}(w,w)&=\frac{2N(C,w)}{4w-3}.
\end{align*}
The first inequality says that $H$ is strictly concave in the charged net cost. To determine the direction of improvement at the boundary $x_1=w$, note that $\partial N/\partial w=16C(1-4C+4w)>0$ and hence
\[
N(C,w)\ge N(C,0)=(4C-1)(8C-3)>0.
\]
Since $w\le C<1/4$, the denominator $4w-3$ is negative. Thus $H_{x_1}(w,w)<0$. Strict concavity makes the derivative remain negative throughout $x_1>w$: inducing a first-price mixture only lowers revenue relative to its boundary value.

At that boundary, $H(w,w)=G(w)$, so the unique best choice for each $w$ is $x_1=w$. Strict increase of $G$ then forces $w=C$. Substituting in the unique later-payment solution from Step 2 gives
\[
(x_1,x_2,x_3)=(C,C,0),
\]
and the recursion yields $(\sigma_1,\sigma_2,\sigma_3)=(1,1-2C,1)$ and $R_1=1+(1-2C)^3$.

\paragraph{Step 4: verify sorting and choose the opening payment.}
The war optimization has so far treated two high signals as revealed. We now verify that a sorting opening supports every war vector used in the comparison. For any such vector, let $L_1$ be the normalized optimal continuation payoff of a low type that unexpectedly stayed at the opening and now faces a high rival at $a/2$.

The low type's true value is $a$. From price $a$ onward, allocation gives nonpositive gross surplus, and net stay costs are nonnegative, so its continuation payoff is at most zero. At $a/2$, exit pays $(1-\sigma_1)/4$, while stay pays at most $-x_1+(1-\sigma_1)/2$. If $\sigma_1=1$, its best payoff is zero. Otherwise high--high indifference gives
\[
x_1=(1-\sigma_1)3/4+\sigma_1W_2
\ge(1-\sigma_1)3/4.
\]
The bound makes staying worse for the low type than exiting. Its optimal continuation payoff is therefore
\[
0\le L_1=(1-\sigma_1)/4\le x_1/3\le C/3.
\]

A low type receives zero by exiting at the opening and $-x_0+qL_1$ by entering. A high type gains $1/2$ against a low rival, an event of probability $1-q$, and enters a war worth $W_1$ against a high rival, an event of probability $q$. Subtracting $x_0$ gives the two opening conditions:
\begin{equation}
x_0\ge qL_1,\qquad
x_0\le (1-q)/2+qW_1.
\label{eq:opening-conditions}
\end{equation}
The first prevents low types from imitating high types. The second ensures that high types prefer entering to the opening lottery. Choosing $x_0=C$, or equivalently $\rho_0=0$, satisfies both strictly because $C<(1-q)/2$ and $qL_1\le qC/3<C$. Assigning every opening stayer a high rival signal, and prescribing sequential best responses for off-path low types, completes a sorting PBE for every war vector.

For any opening payment that supports sorting,
\[
\Pi(\rho)=aq^2R_1-2q\rho_0.
\]
The first term is the war revenue weighted by the probability of two high signals. The second is the expected opening transfer, since the expected number of high opening stayers is $2q$. The war is unaffected by a sunk opening payment, so zero opening reimbursement strictly dominates every positive feasible opening payment. Combining this conclusion with the unique war optimum proves the policy claim and \eqref{eq:backload-rev}.

Finally, subtracting the benchmark formulas gives \eqref{eq:gap-pres}--\eqref{eq:gap-full}. The preservation and full-reimbursement differences are positive immediately. For the difference from None, put $f(x)=9-42x+92x^2-64x^3$. On $[0,1/4]$, $f''(x)=184-384x>0$ and $f'(1/4)=-8$, so $f$ decreases throughout that interval and $f(x)\ge f(1/4)=13/4>0$. This proves all three strict comparisons.
\end{proof}

The optimum in this proof uses the value of later play to support an earlier unpaid action. The final payment changes the middle-price mixture, and that mixture changes the willingness to stay at the first war price. The frontier comparison keeps track of this dependence across rounds while evaluating every feasible payment vector.

\subsection{More bidders and richer signals}
\label{sec:extensions}

The timing theorem optimizes revenue by changing behavior in a particular binary, two-bidder game. We now return to the preservation question: given frictionless behavior that the seller wants to sustain, where must its payments go? The next two examples vary the bidder count and signal space. They illustrate how the opening margin and later information determine the cheapest preserving schedule. 
%they do not claim a revenue optimum for these larger environments.

\subsubsection{The opening margin with more bidders}

Consider $n\ge2$ bidders with independent binary signals, each high with probability $q\in(0,1)$. Let $M$ be the number of high signals, and suppose the common value is $v(M)$, where $v$ is strictly increasing. Set the opening price to $p_0=v(0)$ and the price increment to $\Delta>0$. Assume that every possible value lies on the price grid and is no greater than the cap:
\[
v(m)-v(0)\in\Delta\mathbb Z_{\ge0}\quad(0\le m\le n),
\qquad p_T\ge v(n).
\]
Each stay costs $c>0$.

In the frictionless sorting target, low types exit at the opening and high types stay. If at least two remain, their opening actions reveal $M$. They then stay below $v(M)$ and exit together there, or reach the cap if $v(M)=p_T$. Thus competition among revealed high types leaves them zero surplus.

A high type nevertheless has a positive opening margin. Consider the event that every rival is low. Staying wins the object for certain at $v(0)$, with surplus $v(1)-v(0)$. Exiting instead joins the opening all-exit lottery and receives a $1/n$ share of that surplus in expectation. If any rival is high, entering the subsequent competition gives zero surplus, and exiting also gives zero. Weighting the sole-high event therefore gives
\begin{equation}
w_0=(1-q)^{n-1}\left(1-\frac1n\right)\bigl(v(1)-v(0)\bigr).
\label{eq:nmargin}
\end{equation}
The three factors are the probability that every rival is low, the gain in allocation probability from a certain win instead of a lottery share, and the available surplus.

Every target-path post-opening stay has zero margin. Proposition~\ref{prop:zero} therefore forces payment $c$ at those stays. At the opening, the high type can bear net cost up to $w_0$, so the minimum payment is $\max\{0,c-w_0\}$. Appendix~\ref{app:extensions} specifies the beliefs and checks optimality for every possible observation. For that completion, these are the unique least-cost payment amounts at positive-probability target stays.

The expected number of opening stays is $nq$. Relative to full reimbursement, the seller consequently saves $nq\min\{c,w_0\}$ while preserving allocation and gross receipts. The distinction between preservation and strict implementation matters when $c>w_0$: the minimum opening payment makes high types indifferent. It sustains the target but does not make that target the unique or strict outcome.

\subsubsection{Why a preserving schedule can be nonmonotone}

With more signal categories, a later price can reveal additional information and create a fresh positive incentive margin. Consider two independent signals in $\{0,2,4\}$, with probabilities $(\eta_0,\eta_M,\eta_H)$ that are all positive and sum to one. Here the subscripts distinguish the zero, middle, and high signals. The common value is the sum of the two signals. Decision prices are $0,\ldots,7$, the cap is $8$, and each stay costs $c>0$.

The target separates types twice. Signal zero exits at price $0$, while both positive types stay at $0,1,2,3$. At price $4$, signal two exits and signal four stays. If both bidders remain after that price, both signals must be four, and they stay at $5,6,7$ before reaching the cap. Define
$\theta_M=\eta_M/(\eta_M+\eta_H)$, the probability of a middle signal conditional on a positive signal. The unique least-cost payments at positive-probability target stays are
\begin{center}
\begin{tabular}{@{}cl@{}}
\toprule
Price(s) & Reimbursement per stay\\
\midrule
$0$ & $\max\{0,c-\eta_0\}$\\
$1,2,3$ & $c$\\
$4$ & $\max\{0,c-\theta_M\}$\\
$5,6,7$ & $c$\\
\bottomrule
\end{tabular}
\end{center}

The two partial-payment opportunities come from separation. At the opening, a signal-two bidder gains one unit by replacing a half share of the two-unit surplus with a certain win when its rival has signal zero. This event has probability $\eta_0$, giving opening slack $\eta_0$. At price $4$, a signal-four bidder facing signal two has value $6$ and a two-unit surplus. Staying secures that surplus, whereas joining the all-exit lottery gives half of it. The gain is again one unit, now weighted by the conditional probability $\theta_M$. These margins are the amounts that can be left unreimbursed at the two separating prices.

Between the separating prices, the weaker positive type has zero-margin stays and forces full payment at $1,2,3$. Those public payments also reach the stronger type. After price $4$, two remaining signal-four bidders again have zero-margin continuation, forcing full payment at $5,6,7$. The resulting schedule can move from partial to full, back to partial or zero, and then to full reimbursement. Appendix~\ref{app:extensions} verifies the incentive and belief calculations. The example shows how information and incentive margins, rather than the chronological order of prices alone, determine the preserving payments.

\subsubsection{The role of discrete signals and a finite price grid}

Both examples use a discrete signal space with positive slack at separating decisions. At a continuous indifferent signal cutoff, staying types can lie arbitrarily close to indifference. A public net cost must then be affordable even for these types, forcing exact reimbursement at that history under the assumptions in Appendix~\ref{app:limits}.

Refining the clock raises a different issue. With a fixed physical cost for every stay, a finer grid creates more potentially costly decisions. In a two-bidder continuation with known common value, the total affordable number of advances is bounded. Starting at the first positive tick, unreimbursed receipts therefore tend to zero as ticks shrink, while preserving every frictionless stay becomes arbitrarily expensive. If the cost instead shrinks proportionally with the tick size, the symmetric continuation has a nondegenerate stopping-price limit. Appendix~\ref{app:limits} separates these two scalings and gives their proofs. They describe the limits of the model's discrete margins and cost specification, rather than extending the timing theorem to every finer clock.

\section{Learning experiments}
\label{sec:experiments}

The theoretical analysis specifies how bidders can play in equilibrium under a reimbursement rule. We now ask whether bidders who learn from repeated play recover that behavior, and whether the seller earns the predicted revenue. The first experiment tests the analytical comparisons with two, three, and four bidders. The second searches over payment schedules in a four-bidder game outside the scope of the two-bidder optimality theorem.

For the first experiment, we call each policy's analytical prediction its \emph{reference profile}. A profile specifies a strategy for every bidder. Preserve and Full share the frictionless target of Section~\ref{sec:preservation}; None and back-loading have different continuation strategies. This terminology distinguishes a prediction about play under a particular policy from the frictionless assessment that a preserving rule is designed to sustain.

Code and saved results for both experiments are available at \url{https://github.com/mirmnel/reimbursements}. The archive's README gives commands to rerun the experiments, print the summaries reported here, and redraw Figure~\ref{emp:fig-search-ranked}. The training description below records the seed and initialization choices.

\subsection{Testing the analytical predictions}

\paragraph{From an auction rule to a learned profile.}
For each learning problem, the experiment separates prediction from training and evaluation:
\begin{enumerate}[leftmargin=*,itemsep=3pt,topsep=3pt]
    \item \emph{Build the auction.} Fix the number of bidders, signals, price grid, stay cost, and reimbursement rule.
    \item \emph{Compute the reference.} Use sorting at the opening and backward induction to determine the later stay probabilities.
    \item \emph{Check the reference.} Evaluate exact best responses, including opening deviations, and check that the reference has zero ex ante deviation gain up to numerical precision.
    \item \emph{Train the bidders.} Let them learn through repeated play from a given random seed. The learners never receive the reference profile.
    \item \emph{Evaluate learned play.} Score the final average strategy for unilateral deviation gains, agreement with the reference, and expected seller net revenue.
\end{enumerate}
This order makes the comparison meaningful: the analytical prediction is checked before learning, and the learned strategy is then evaluated independently of whether it resembles that prediction. As explained below, Full's learned profiles can be reused across costs, so the number of scored profiles need not equal the number of distinct training runs.

\paragraph{Auction settings and payment rules.}
We use the binary wallet game with $n\in\{2,3,4\}$ bidders, $a=1$, and $q=0.3$. The common value is the sum of the signals: $v(M)=Ma$ when $M$ signals are high. The decision prices are $0,1/2,\ldots,n-1/2$, followed by terminal cap $n$. The tested costs are $c=\kappa\in\{0.025,0.05,0.1\}$. Since $a=1$, monetary cost $c$ and normalized cost $\kappa$ have the same numerical value.

Every tested cost lies below $1/4$ and below the high type's opening margin $w_0$ in \eqref{eq:nmargin}. That margin is $0.350$, $0.327$, and $0.257$ at the three bidder counts. Thus the tested high types can bear the opening cost while preserving the sorting target. The choice of $q$ also keeps revealed-value wars sufficiently frequent to observe the revenue differences. Increasing $q$ reduces the opening margin: with four bidders it reaches the largest tested cost, $0.1$, when $q=1-(2/15)^{1/3}\approx0.49$. Reducing $q$ makes the events with at least two high bidders less frequent.

At every setting we compare three anonymous round-only rules, each paying every stayer the same amount in a given round:
\begin{itemize}[leftmargin=*,itemsep=2pt,topsep=3pt]
    \item None, $(0,\ldots,0)$, leaves every stay unpaid.
    \item Preserve, $(0,c,\ldots,c)$, fully reimburses post-opening stays and leaves the opening unpaid because $c<w_0$.
    \item Full, $(c,\ldots,c)$, reimburses every stay, including the opening.
\end{itemize}
With two bidders we also test back-loading, $(0,0,0,c)$, from Theorem~\ref{thm:backload}. Five seeds per setting give 150 scored cases from 120 distinct training runs. For each bidder count and seed, Full is trained once in the frictionless game, and the resulting strategy is evaluated at all three costs. Full reimbursement gives bidders their frictionless payoff on every path, so changing $c$ changes the seller’s expenditures without changing the bidders’ learning problem.

\paragraph{What the reference predicts.}
In every reference profile, low types exit and high types stay at the opening. If at least two stay, their opening actions reveal the common value. The rules then differ in what happens during the revealed-value war.

With two high stayers, None uses the probabilities in \eqref{eq:none-strategy}. In the two-bidder game, back-loading uses the probabilities in Theorem~\ref{thm:backload}. Under Preserve and Full, active high types stay certainly below the revealed value. With three or more high stayers, the unreimbursed continuation is computed by backward induction. In these war continuations, the reference selects certain staying whenever it is a best response.

These predictions give two revenue comparisons. Preserve should save opening transfers worth $nq\kappa$ relative to Full while supporting the same play. At two bidders, back-loading should earn the most of the four tested rules and exceed None by the amount in \eqref{eq:gap-none}.

\paragraph{How bidders learn.}
We use external-sampling Monte Carlo counterfactual regret minimization (MCCFR) \citep{zinkevich2007regret,lanctot2009montecarlo}, with regret matching, random initial regrets, and linearly weighted average strategies \citep{brown2019discounted}. Each run has 2,000,000 iterations, or 100,000 in the schedule search, which trains 6,260 runs on a clock with four decision prices instead of up to eight. We evaluate the weighted average of the strategies played during training.

Training uses seeds $0$--$4$ for the prediction tests and $0$--$9$ for the search. With $n$ bidders, initial regrets are drawn uniformly from $[0,0.05n]$.

Bidders share one regret table indexed by their own signal and public history after relabeling bidder identities. They consequently learn a symmetric strategy, while retaining the ability to mix between stay and exit. This training choice aligns the experiment with the symmetry of the analytical continuations. Symmetry restricts training only. When we check for profitable deviations, a single bidder may switch to any strategy, symmetric or not.

\paragraph{What the scores measure.}
For a fixed learned profile, expected revenue and deviation gains are evaluated exactly, averaging over signals, mixed actions, and terminal lotteries. The \emph{utility loss} is the largest ex ante gain any bidder could obtain by replacing its strategy with a best response against the others' fixed strategies. Best-response evaluation is also central to \citet{deon2024iterative}.

We use two numerical criteria, both with thresholds we chose. A profile passes the equilibrium criterion when its utility loss is at most $10^{-3}n/2$. The factor $n/2$ scales the threshold with payoffs, since the common value sums $n$ signals. A passing profile \emph{matches the reference} when its stay probability differs from the reference by at most $0.05$ for each bidder and each observation that occurs with positive probability under the reference profile. The first criterion asks how much an individual bidder could gain by changing its behavior. The second asks whether learned play resembles the analytical continuation being tested. The theory does not determine either threshold, and revenue enters neither. All learned utility losses are well below the equilibrium threshold (Table~\ref{tab:learned-profiles}).

An ex ante score averages over how often histories are reached. A small loss can therefore coexist with a larger conditional error at a rare history. These checks do not establish sequential optimality for every possible observation, and regret minimization has no general convergence-to-Nash guarantee for the auctions studied here.

\paragraph{Learned strategies.}
Table~\ref{tab:learned-profiles} summarizes the 150 scored profiles. All pass the numerical equilibrium criterion, and 141 match their reference profile. Every two- and three-bidder profile matches, with maximum stay-probability differences $0.002$ and $0.008$, respectively. Every four-bidder profile under Preserve and Full also matches.

\begin{table}[htbp]
\centering
\caption{Learned profiles by bidder count. Columns after Matched report maxima over profiles of the stay-probability difference from the reference, utility loss, and revenue error from theory.}
\label{tab:learned-profiles}
\begin{tabular}{@{}rcccc@{}}
\toprule
$n$ & Matched & Stay diff. & Utility loss & Rev.\ error\\
\midrule
$2$ & $60/60$ & $0.002$ & $4.7\cdot10^{-5}$ & $0.0004$\\
$3$ & $45/45$ & $0.008$ & $1.1\cdot10^{-4}$ & $0.0004$\\
$4$ & $36/45$ & $0.12$ & $2.5\cdot10^{-4}$ & $0.0011$\\
\bottomrule
\end{tabular}
\end{table}

Every mismatch occurs with four bidders under None. Six of those 15 profiles match, all at $\kappa\le0.05$; the other nine differ by up to $0.12$ in stay probability while retaining utility losses at most $2.5\cdot10^{-4}$. Most war decisions in the reference are mixed, so a bidder is indifferent against the reference rival mixture. Together with the low reach probability of some histories, this helps explain why noticeable strategy differences can coexist with small ex ante loss. In every run, learned revenue is within $0.0011$ of theory.

\paragraph{Learned revenue.}
In every setting, Preserve minus Full matches $nq\kappa$ to four decimal places. Since the rules predict identical play and differ only in opening transfers, this comparison chiefly checks that learned bidders take the predicted opening actions.

Back-loading gives a more demanding comparison because it changes continuation behavior. At two bidders, its gain over None, averaged over seeds, is within $1\%$ of \eqref{eq:gap-none} at each cost. It earns the most of the four rules in all 15 seed--cost pairs. Thus the learned revenue comparison recovers the benefit of changing which continuation decisions are reimbursed.

\subsection{Searching over seller schedules}

The model admits any reimbursement schedule, but our theory ranks schedules only in the two-bidder, low-cost case. Searching over schedules with learned play extends the design beyond that case. We use four bidders with $q=0.3$, $a=1$, decision prices $0,1,2,3$, and cap $4$. The cost is $c=w_0+0.05\approx0.307$, above the opening margin. Preserving the frictionless sorting target now requires a positive opening payment. A revenue-oriented policy can instead change the continuation and leave that opening unpaid.

\paragraph{Search and paired comparisons.}
Each entry of $\rho=(\rho_0,\rho_1,\rho_2,\rho_3)$ takes a value in $\{0,c/4,c/2,3c/4,c\}$, giving 625 grid schedules. These include None, Full, back-loading, and war-only reimbursement $(0,c,c,c)$. The least-cost preserving rule $(c-w_0,c,c,c)$ lies off the grid and is evaluated separately. The distinction matters when interpreting the figure: the ranked curve contains the 625 grid schedules, with preservation added as a comparison point at its insertion rank.

The search retains learning and exact evaluation but omits reference construction and matching, since most schedules have no analytical reference. Each rule uses the same ten seeds. We rank schedules by mean revenue. A paired comparison first subtracts their revenues within each seed, then reports the mean and standard error of those ten differences. All $625\cdot10+10=6,260$ runs pass the numerical equilibrium criterion.

\begin{figure}[htbp]
\centering
\includegraphics[width=0.92\linewidth]{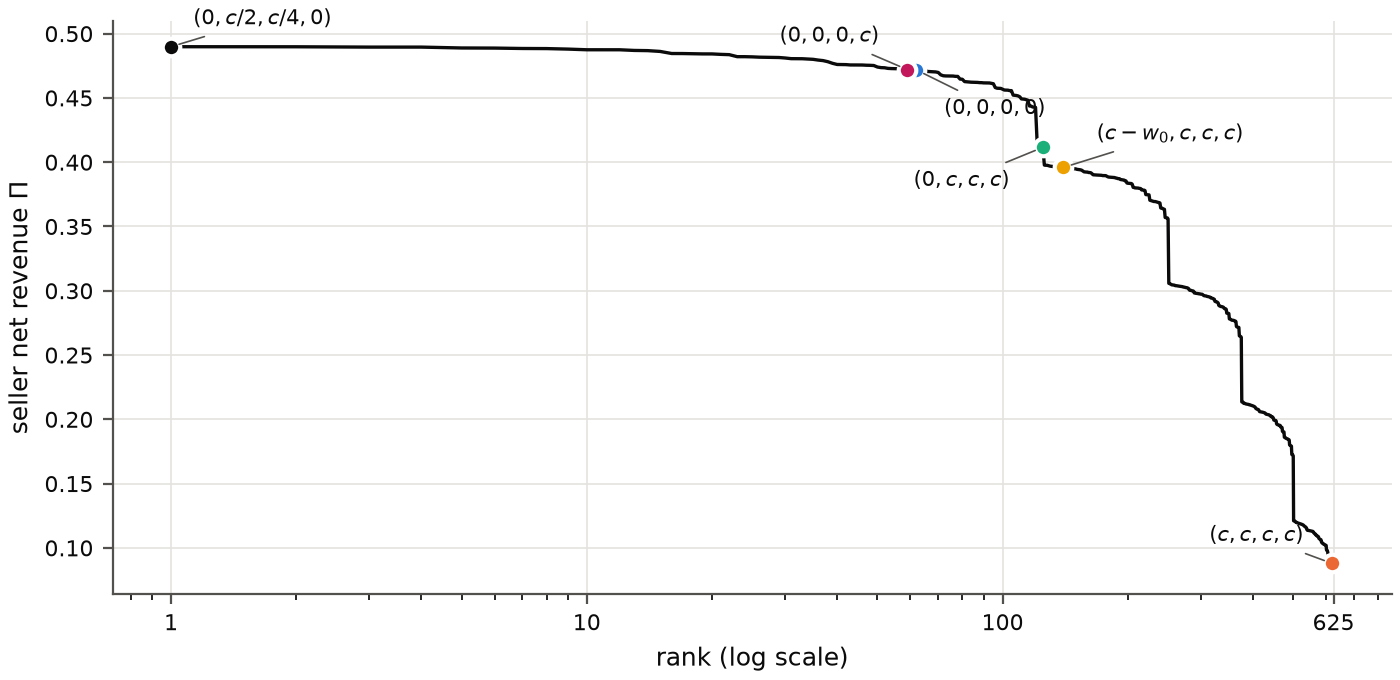}
\caption{The 625 grid schedules ranked by learned net revenue, with rank on a log scale. Markers identify the best tested schedule and named rules. The off-grid least-cost preserving rule $(c-w_0,c,c,c)$ is shown at its insertion rank by revenue; it is not one of the 625 grid schedules.}
% Figure description: Seller net revenue falls from about 0.49 to 0.09 across 625 schedules. The best schedule pays nothing at the opening, half the cost at price one, a quarter at price two, and nothing at price three. Back-loading and no reimbursement are close to 0.472; preservation and full reimbursement earn less.
\label{emp:fig-search-ranked}
\end{figure}

\paragraph{Search results.}
Figure~\ref{emp:fig-search-ranked} shows that the best tested grid schedule at learned play is $(0,c/2,c/4,0)$, with mean net revenue $0.4898$. Its advantage over the runner-up, $(0,c/2,c/4,c/4)$, is $0.0001\pm0.0002$, so the two are effectively tied at the reported precision. All ten highest-ranked schedules pay nothing at the opening and $c/2$ at price $1$.

The best schedule exceeds None ($0.4718$) by $0.0180\pm0.0001$ over paired seeds, about $4\%$. Back-loading earns $0.4720$, only $0.0002\pm0.0003$ more than None. War-only reimbursement earns $0.4116$, least-cost preservation earns $0.3959$, and Full earns $0.0879$. The successful schedule therefore places a partial payment earlier than the last stay and outperforms each named baseline in this experiment.

\paragraph{Economic interpretation and scope.}
Holding sorting behavior fixed, each additional $c/4$ paid at the opening costs $nq\,c/4\approx0.09$. An opening payment must earn back this expenditure through its effect on behavior if it is to benefit the seller. Here the reported search favors leaving the opening unpaid and using a partial payment to support later competition. Paying the full cost at price $1$ instead adds more in reimbursements than in sale receipts.

This finding complements the exact timing theorem: the preferred payments depend on the game being financed. The search changes the bidder count, cost region, and clock relative to Theorem~\ref{thm:backload}. It establishes a profitable tested alternative, rather than an unrestricted optimum or a causal effect of cost alone. The reported standard errors describe variation across training seeds, not uncertainty across a population of auction instances.

\FloatBarrier

\section{On-chain realization and bidder protection}
\label{sec:implementation}

The economic analysis assumes that a choice is recorded and that a promised transfer is paid. An on-chain realization must give those events the same meaning as in the model. In particular, accepted stays must authorize the intended future prices, simultaneous choices must remain hidden until a common release, and reimbursement commitments must be credible before bidders act. We first show what affirmative continuation guarantees about purchase-price exposure, then discuss the further conditions needed to implement the game.

\subsection{Fresh authorization and purchase-price exposure}

Timely inclusion is economically different from eventual inclusion. Consider an exit-only clock in which an admitted bidder remains active until its exit becomes effective. A bidder with known value 100 may try to exit at 100 yet become the sole remaining bidder at 130 if its message is censored while a rival exits later. A separately enforced personal cap would prevent that outcome; the unqualified default-stay rule does not.

Affirmative continuation makes continued eligibility depend on a new accepted stay. Failure to record that response removes the bidder from future rounds. This protects against unintended later prices at the cost of premature exclusion: a bidder that wanted to remain can lose eligibility when its stay is missing or late.

To state the price bound, let $P_i(z):=X_i(z)P(z)$ be bidder $i$'s purchase-price debit, equal to zero if it loses. Let $t_i^{\mathrm{last}}\in\{-1,0,\ldots,T-1\}$ be its last round with an accepted stay, using $-1$ if it never stays. Suppose only authenticated in-round responses are accepted, a missing stay means exit, and eligibility is never restored.

\begin{proposition}[Purchase-price exposure]
\label{prop:exposure}
On every terminal path, bidder $i$'s purchase-price debit satisfies
\begin{equation}
P_i(z)\le p_{t_i^{\mathrm{last}}+1},
\label{eq:exposure}
\end{equation}
where the right side is $p_0$ when $t_i^{\mathrm{last}}=-1$. A bidder without an accepted stay at $p_t$ cannot receive the object later, although it may still receive it at $p_t$ through the all-exit lottery.
\end{proposition}

\begin{proof}
If $i$ has no accepted stay at $p_t$, it either receives the object in the current all-exit lottery or exits permanently. Thus a winning bidder allocated at $p_t$ must have stayed in the preceding round, unless $t=0$; allocation at the cap requires a stay at $p_{T-1}$. A loser has zero debit and the prices are nonnegative. These cases give \eqref{eq:exposure}.
\end{proof}

The one-step term follows from the settlement convention. A stay at $p_t$ authorizes entry into the round at $p_{t+1}$. If everyone then exits, the all-exit lottery can allocate the object at $p_{t+1}$ without another stay. The proposition consequently bounds the purchase price authorized by past participation; it does not bound every loss a bidder can suffer.

Censorship may exclude a bidder that wanted to continue, strand earlier costs, and reduce competition. One extension is to ask whether reimbursement or carefully bounded over-reimbursement can insure bidders against perceived exclusion risk. Such insurance lies outside the present no-overpayment model and cannot make a censored response count.

\subsection{Hidden choices and credible commitments}

An accepted response must be bound to the auction, bidder, round, and price, and stale responses must be rejected. Otherwise a response intended for one decision could be used to authorize another. These requirements give operational content to the proposition's assumptions about authenticated in-round actions and irreversible exit.

The economic model also uses simultaneous choices. A realization must therefore hide current responses until a common release and make those choices binding. Ordinary commit--reveal hides a choice but permits strategic non-reveal after observing others' openings, so it requires forced opening or an incentive-compatible penalty. These are protocol obligations, not consequences of the equilibrium analysis \citep{galal2018verifiable,alpos2026censorship}.

The reimbursement rate for a round must be fixed using public information available before that round's hidden choices. An immutable policy address is insufficient if its code or dependencies can change. The implementation must also fund both the promised transfers and delivery of the auctioned asset. A commitment to an amount that cannot be paid would change the continuation incentives analyzed in the paper.

\subsection{Funding and the timing of participation costs}

Expected expenditure and required collateral answer different questions. The objective in \eqref{lp:objective} averages transfer payments under the target assessment. A fully prefunded implementation instead needs a balance covering payments on every executable path. This balance can be much larger than expected expenditure, and our economic model assigns no opportunity cost to the locked capital.

The physical cost must also be attached to the modeled action on every branch. In the model, a stay incurs its cost and earns its reimbursement even when it immediately wins. In a commit--reveal implementation, both stayers and exiters may pay for their current messages, while a later message is never sent if the auction has already ended. Any current cost common to both actions cancels in their local comparison; a stay can additionally create a future response obligation by keeping the auction going.

These branch distinctions determine whether a proposed fee and transfer system implements the model. Crediting a stayer immediately, or only after the next response, gives the modeled payoff only if terminal accounting also matches the branches on which no next response occurs. %The implementation claim is therefore a specification of required semantics, not a validation of a deployed contract.

\section{Discussion and conclusion}
\label{sec:conclusion}

Recurring participation costs make the financing of continued eligibility part of auction design. We have separated the question of sustaining a specified frictionless equilibrium from the question of earning the most net revenue under changed behavior. For a fixed finite target, the first question has a transparent LP formulation. Zero incentive margins explain why some stays require exact reimbursement. Positive margins identify cost that bidders can bear, while public payments create expenditure on stronger types as well as the weakest type whose incentives require them.

The binary wallet example shows why the second question can have a different answer. Within the stated low-cost symmetric sorting comparison, the unique optimum reimburses the last preterminal stay and leaves the earlier stays unpaid. That final payment supports a continuation payoff that affects earlier decisions. Some auctions end sooner than under frictionless play, reducing gross sale receipts, but the seller saves more in transfers. The policy's advantage comes from the behavior induced by the schedule and its net accounting, rather than from changing the timing of transfers for fixed behavior.

The learning experiments test both the analytical revenue comparisons and the strategies behind them. Most learned profiles match their policy-specific reference closely, and revenue differences remain accurate even in the four-bidder unreimbursed profiles that differ more in their action probabilities. The exact best-response checks and the reference-matching criterion measure different aspects of these outcomes. The schedule search shows how to design reimbursements where the theory gives no optimum. Learning with exact evaluation finds a schedule with partial payments in the middle rounds that earns more than the named rules. Together, the experiments support treating payment timing as an environment-dependent design choice; they do not extend the two-bidder theorem to every bidder count, cost region, or clock.

Several boundaries remain explicit. The optimality result selects a symmetric sorting branch and uses a particular allocation convention. Preservation need not select a unique equilibrium, and the learning procedure does not test every possible equilibrium-selection process. Continuous signal cutoffs can eliminate the positive slack available with discrete types. A finer price grid also changes the problem unless the cost per stay is scaled with it. The supplementary calculations make these distinctions precise.

Finally, an on-chain deployment must realize the decisions and transfers that the economic model assumes. Affirmative continuation bounds purchase-price exposure after the last accepted stay, while timely inclusion, hidden and binding choices, collateral, and settlement require their own protocol guarantees. Within these conditions, the design lesson is to identify which continuation decisions are worth financing and to compare the resulting sale receipts with the full reimbursement bill.

\section*{Acknowledgments}
Edwin Lock and Francisco Marmolejo-Coss\'{\i}o were supported
by a research grant from IOG for this work.

\section*{Use of generative AI}
We used Anthropic's Claude through Fable, OpenAI's GPT-5.6 through ChatGPT, and Codex to assist with theoretical exploration, calculations, coding, and manuscript preparation. The authors are responsible for all research claims and the final manuscript.

\bibliographystyle{plainnat}
\bibliography{refs}

\clearpage
\appendix
\section{Supplementary preservation and extension arguments}
\label{app:supplementary}

The main text proves the preservation program, the zero-margin condition, and the binary revenue-design theorem. This appendix first records the round-only specialization of the preservation program, then explains the optional local-deviation reduction and verifies the beliefs and incentives used in the larger preservation examples. As throughout the paper, continuation payoffs exclude costs already incurred.

\subsection{Round-only policies}

The public-history formulation allows a different payment at every history. To study payment timing by round, suppose physical costs also depend only on round: $c_i(h)=c_t$. Write $\delta_t=c_t-\rho_t$ for unreimbursed cost in round $t=0,\ldots,T-1$. The seller now chooses a vector of net costs, one entry per round.

Each incentive comparison in \eqref{lp:ic}, indexed by $j=(i,\iota,\tau_i)$, has frictionless margin $m_j=U_i^0(\sigma_i^0\mid\iota)-U_i^0(\tau_i\mid\iota)\ge0$. Let $d_{jt}$ be the target's conditional probability of an own stay in round $t$ minus the deviation's, given the initial observation $\iota$. Rounds before the decision at $\iota$ have coefficient zero. Apply \eqref{eq:affine} to both plans and subtract. The frictionless terms give $m_j$, while their different stay probabilities give $d_j\cdot\delta$. Thus the target's costly payoff advantage is $m_j-d_j\cdot\delta$.

These exposure coefficients can be negative. If a deviation makes more stays in some round, leaving that round costly discourages the deviation rather than the target. The incentive constraint must retain that sign; it cannot replace all stay differences with nonnegative cost burdens.

For the expenditure calculation, let $\Lambda_t$ be the target's expected total number of stays in round $t$. Reducing the round's reimbursement by $\delta_t$ saves $\Lambda_t\delta_t$ because the target behavior is fixed. With $c,\delta,\Lambda$, and $d_j$ denoting the corresponding round vectors, the maximum saving relative to full reimbursement is
\begin{equation}
\max_{0\le\delta\le c}\ \Lambda\cdot\delta
\quad\text{subject to}\quad d_j\cdot\delta\le m_j\quad\forall j.
\label{eq:roundlp}
\end{equation}
Here incentive exposure $d_j$ determines which costs can be left with bidders, while payment traffic $\Lambda$ determines how much the seller saves. These are different roles: a frequently paid round is attractive for savings only if the incentive constraints permit a reduction there.

Equation~\eqref{eq:roundlp} also clarifies why a single reimbursement fraction can be wasteful. A common unreimbursed fraction restricts $\delta$ to $\delta=\eta c$, where $\eta\in[0,1]$. Separate rates allow the seller to place more of the net burden in rounds whose incentive constraints permit it, while reimbursing more tightly constrained rounds. Section~\ref{sec:extensions} gives examples with partial and nonmonotone payments.

\subsection{When local deviation checks suffice}
\label{app:preservation}

The preservation LP in Section~\ref{sec:preservation} compares complete continuation plans. This appendix gives the belief condition under which one can instead check each current action followed by a return to the target. The condition matters for observations that can occur under a deviation. Evaluating a later decision must use the same posterior that the bidder would obtain by updating her earlier belief through the deviation.

Call beliefs \emph{deviation-consistent} if, whenever a bidder's later observation has positive probability under one of her unilateral continuation deviations, her assigned belief given that observation agrees with updating her current belief through the deviation and her rivals' target play. Under this condition, the following backward replacement argument justifies the reduction.

\paragraph{Backward replacement.}
Suppose beliefs are deviation-consistent and the target weakly beats every current action followed by the target continuation for every possible observation. Fix a bidder and a pure continuation deviation starting from an arbitrary observation. At the last decision date, replace each action prescribed by the deviation with the target action or mixture. For each observation that can occur at that date under the deviation, deviation consistency identifies the posterior with the belief used by the local inequality. The replacement therefore cannot lower the bidder's expected payoff from her initial observation. Work backward until every action has been replaced. The target therefore weakly beats the original complete plan. Randomized plans follow by averaging; necessity is immediate because a one-action deviation is a complete plan. The argument also proves the sufficient $w/D$ bound in the main text: target stays bear at most total net cost $w$, while adding nonnegative costs cannot make a target-exiting type's stay deviation more attractive.

\subsection{Extension calculations}
\label{app:extensions}

\subsubsection{Binary signals with any finite number of bidders}

Section~\ref{sec:extensions} calculated the opening margin on the sorting path. We now specify beliefs at later histories and check the continuation decisions needed to complete the assessment.

\paragraph{Beliefs after the opening.}
Let $S_{\mathrm{open}}$ be the original set of opening stayers. At later histories, each active bidder believes that the other members of $S_{\mathrm{open}}$ are high and all opening exiters are low; these beliefs retain the original opening information after unexpected later exits. A high type then believes the value is $v(|S_{\mathrm{open}}|)$ and stays below that price, exiting at or above it. A low type that unexpectedly stayed at the opening believes the value is $v(|S_{\mathrm{open}}|-1)$ and exits.

\paragraph{Continuation decisions.}
In the frictionless game, a high bidder's continuation below its believed value yields zero surplus: all rivals stay until that value. Earlier exit also gives zero. At or above the believed value, all rivals exit; joining the forced tie weakly beats becoming the sole stayer because surplus is nonpositive. An off-path low bidder cannot gain by following rivals to their higher exit price; at a forced tie above its own value, sharing the negative surplus is preferable to becoming the sole stayer. This verifies sequential rationality at continuing histories.

\paragraph{Opening incentives and minimum payments.}
At the opening, a high type has a positive gain only when every rival is low. Staying then obtains the full surplus $v(1)-v(0)$, while exiting obtains a $1/n$ share, giving \eqref{eq:nmargin}. A low type cannot obtain positive surplus by staying: with no high rivals its value equals the opening price, and with high rivals it makes them infer a value above the true one. This completes the frictionless target.

Every target-path post-opening stay has zero margin, so Proposition~\ref{prop:zero} forces exact reimbursement there. Fully reimburse all post-opening histories to retain the specified off-path continuation as well. The high type's opening constraint becomes $w_0-(c-\rho_0)\ge0$; the low type's deviation acquires only a nonnegative net cost. Thus the minimum is $\rho_0=\max\{0,c-w_0\}$. Every opening payment and every required later payment has positive objective weight, proving uniqueness on target-path stays. Full reimbursement pays an extra $nq\min\{c,w_0\}$ at the opening and the same amount later.

\subsubsection{The three-signal wallet}

The two separating prices, $0$ and $4$, supply the possible net-cost budgets. The remaining checks establish that using those budgets does not create another profitable deviation.

\paragraph{Posteriors and unexpected own actions.}
At prices $1,\ldots,4$, use the rival's prior conditional on being positive; at prices $5,\ldots,7$, use rival signal four. Apply these same posteriors to own types whose earlier actions were unexpected. Signal zero exits after an unexpected opening stay, and signal two exits after unexpectedly passing price $4$.

\paragraph{Necessary payments.}
Stays at $1,2,3$ by signal two and at $5,6,7$ by signal four have zero frictionless margins, forcing full reimbursement. Allow net costs only at the separating prices, writing $k_0=c-\rho_0$ and $k_4=c-\rho_4$. A signal-four type's price-$4$ stay-minus-exit payoff is $\theta_M-k_4$, while a signal-two type's opening margin is $\eta_0-k_0$. Necessity therefore gives
\[
k_0\le\min\{c,\eta_0\},\qquad
k_4\le\min\{c,\theta_M\}.
\]
\paragraph{Incentives on the sorting path.}
For sufficiency, use equality in both bounds and exact reimbursement elsewhere. At prices $1,2,3$, a signal-four type's continuation payoff is $2\theta_M-k_4\ge\theta_M>0$, while signal two is indifferent. At price $4$, staying cannot give signal two positive surplus, whereas signal four's constraint holds by construction. At the opening, signal four's margin before the current cost is
\begin{align*}
2(\eta_0+\eta_M)-(1-\eta_0)k_4
&\ge2\eta_0+\eta_M\\
&>\eta_0\ge k_0,
\end{align*}
so its constraint holds strictly. Signal zero cannot gain by staying against a rival that remains until an overestimated value.

\paragraph{Private off-path cases and optimality.}
For the private off-path cases, signal zero's exit at prices $1,2,3$ yields zero and waiting cannot improve on it. At price $4$ against signal two, its exit payoff is $-1$ and its stay payoff is at most $-2$; against signal four it can exit for zero and cannot gain by waiting. After price $4$, both signal zero and an unexpected signal two can exit for zero against a signal-four rival that stays to $8$. These checks establish optimality for every possible observation under the stated beliefs.

Necessity fixes all full payments and bounds the two remaining net costs; each corresponding target stay has positive probability. Maximizing these net costs therefore uniquely minimizes target expenditure and gives the schedule in Section~\ref{sec:extensions}.

\section{Additional scope observations}
\label{app:limits}

\subsection{Continuous signal cutoffs}

A discrete separating type can have positive incentive slack. At a continuous cutoff, supported types can approach indifference from the staying side. Because a public payment cannot distinguish these nearby private types, their arbitrarily small margins constrain the net cost at the same public history.

Here signals may range over intervals while the clock remains finite. Suppose a target uses an interior cutoff $s^*$ at history $h$, the stay-minus-exit margin $W_i^0(h,s_i)$ is continuous and strictly increasing near $s^*$, and $W_i^0(h,s^*)=0$. Assume there are supported staying types approaching $s^*$ from above for whose observations $(h, s_i)$ optimality is required. Public net cost $k_i(h)$ must be affordable by every such type: the same nonnegativity argument as in Proposition~\ref{prop:zero} gives $0\le k_i(h)\le W_i^0(h,s_i)$. Taking these types down to $s^*$ yields $k_i(h)=0$. This is a local necessity result under the stated cutoff assumptions.

\subsection{A fixed cost as the price grid becomes finer}

This calculation holds the value fixed and adds more decision prices. Its first step is an accounting bound that applies to every PBE, without choosing a symmetric continuation. The key resource is the total surplus available to pay for stays.

Consider two bidders who already know common value $V>0$. For integer $J\ge2$, set $\Delta=V/J$, use decision prices $\Delta,2\Delta,\ldots,V-\Delta$, and retain forced allocation with terminal cap $V$. This continuation begins at $\Delta$ and has no earlier signal-revealing opening. Let $P_\Delta$ be the sale price, $L_\Delta$ the total number of stays, and $u_{i,\Delta}$ bidder $i$'s realized utility. With fixed unreimbursed cost $c>0$, pathwise accounting gives
\[
u_{1,\Delta}+u_{2,\Delta}+P_\Delta+cL_\Delta=V.
\]
Either bidder can exit initially for nonnegative expected utility: it gets zero if its rival stays and a nonnegative share of the surplus if both exit. In every PBE, therefore, $c\E[L_\Delta]\le V$. Each advance requires two stays, so
\[
\E[P_\Delta]\le\Delta\left(1+\frac{\E[L_\Delta]}{2}\right)
\le\Delta\left(1+\frac{V}{2c}\right)\longrightarrow0.
\]
The price bound tends to zero because the expected number of affordable advances stays bounded while the distance advanced at each step shrinks. Preserving frictionless play has the opposite expenditure problem: it finances all $J-1$ decision rounds for both bidders. Every stay on the frictionless all-stay path has zero margin. Preserving that path forces full reimbursement and gives net revenue $V-2c(J-1)\to-\infty$.

\subsection{A cost proportional to the price increment}

We next change the scaling. Reducing the per-stay cost together with the tick size allows competition to survive a nontrivial distance along the clock. Here the calculation concerns the unique symmetric continuation, rather than every PBE.

Instead let $c_\Delta=\gamma\Delta$, with $0<\gamma<1/2$, in the same continuation. The index $j=(V-p)/\Delta$ counts price steps remaining to the cap: $j=1$ is the last decision, and $j$ decreases as the auction advances. This reverses the forward index $t$ used in the binary war. Backward induction gives the stay probabilities of the unique \emph{symmetric} continuation:
\[
\sigma_j=
\begin{cases}
1-2\gamma/j,&j\text{ odd},\\
1,&j\text{ even}.
\end{cases}
\]
Here $W_j$ is a bidder's equilibrium continuation payoff in monetary units, with $W_0=0$ at the terminal cap. The action difference against stay probability $\sigma$ is
\[
(1-\sigma)j\Delta/2+\sigma W_{j-1}-\gamma\Delta.
\]
At odd $j$ the next payoff is zero, giving the stated mixture and $W_j=\gamma\Delta$; at even $j$ the next payoff covers the cost, giving $\sigma_j=1$ and $W_j=0$.
For fixed $p\in(0,V)$, put $m_\Delta(p)=J-\lfloor p/\Delta\rfloor$. To sell above $p$, the auction must pass every decision price at or below $p$. At each such round, both bidders must stay. This gives the survival probability
\[
\Prb(P_\Delta>p)=
\prod_{\substack{m_\Delta(p)\le j\le J-1\\ j\text{ odd}}}
(1-2\gamma/j)^2.
\]
For fixed $p$, $m_\Delta(p)\sim(V-p)/\Delta$, so both endpoints of the product tend to infinity. Only the odd-index rounds contribute factors different from one. Their reciprocal sum is asymptotically half of the corresponding harmonic sum; the squared factors account for the need for two stays. The relevant harmonic sum and error bound are
\begin{align*}
\sum_{\substack{m_\Delta(p)\le j\le J-1\\j\text{ odd}}}\frac1j
&=\frac12\log\frac{J}{m_\Delta(p)}+o(1),\\
\sum_{j=m_\Delta(p)}^{J-1}j^{-2}&=o(1).
\end{align*}
Expanding $2\log(1-2\gamma/j)=-4\gamma/j+O(j^{-2})$ therefore gives survival limit $((V-p)/V)^{2\gamma}$. Since sale prices lie in $[0,V]$, bounded convergence in the tail-integral formula yields
\[
\lim_{\Delta\to0}\E[P_\Delta]
=\int_0^V\left(1-\frac pV\right)^{2\gamma}dp
=\frac{V}{1+2\gamma}.
\]
Preserving the frictionless all-stay path yields
\[
\Pi_\Delta^{\mathrm{pres}}
=V-2\gamma(V-\Delta)\longrightarrow V(1-2\gamma).
\]
Both revenues concern the same known-value continuation and cost scaling. The limiting difference is $4\gamma^2V/(1+2\gamma)>0$, so the unreimbursed symmetric continuation eventually earns more than preservation. 
%This comparison does not rank all equilibria or all reimbursement policies.

\end{document}